\RequirePackage{lineno}
\documentclass[onecolumn,aps,prl,showpacs,superscriptaddress,floatfix,preprint]{revtex4}
\usepackage{epsfig}
\usepackage{color}
\newcommand {\snn}	{\sqrt{s_{_{\rm NN}}}}

\newcommand {\pt}	{p_{\perp}}

\newcommand {\mean}[1]	{\langle #1\rangle}
\newcommand {\Ncoll}	{N_{\rm coll}}

\begin{document}
\title{Charm quarks are more hydrodynamic than light quarks in final-state elliptic flow}
\author{Hanlin Li}
\email{lihl@wust.edu.cn}
\affiliation{Hubei Province Key Laboratory of Systems Science in Metallurgical Process, Wuhan University of Science and Technology, Wuhan 430081,China}
\author{Zi-Wei Lin}
\email{linz@ecu.edu}
\affiliation{Key Laboratory of Quarks and Lepton Physics (MOE) and Institute of Particle Physics, Central China Normal University, Wuhan 430079, China}
\affiliation{Department of Physics, East Carolina University, Greenville, North Carolina 27858, USA}
\author{Fuqiang Wang}
\email{fqwang@purdue.edu}
\affiliation{School of Science, Huzhou University, Huzhou, Zhejiang 313000, China}
\affiliation{Department of Physics and Astronomy, Purdue University, West Lafayette, Indiana 47907, USA}
\date{\today}

\begin{abstract}

We study the charm quark elliptic flow ($v_{2}$) in heavy ion as well as small system collisions by tracking the evolution history 
of quarks of different flavors within a multi-phase transport model. The charm quark $v_{2}$ is studied as a function of the number of collisions the charm quark suffers with other quarks and then compared to the $v_{2}$ of lighter quarks. We find that the common escape mechanism is at work for both the charm and light quark $v_{2}$. However, contrary to the naive expectation, the hydrodynamics-type flow is found to contribute more to the final state charm $v_{2}$ than light quark $v_{2}$. This could be explained by the smaller average deflection angle the heavier charm quark undergoes in each collision, so that heavy quarks need more scatterings to accumulate a significant $v_{2}$, while lighter quarks can more easily change directions with scatterings with their $v_{2}$ coming more from the escape mechanism. Our finding thus suggests that the charm $v_{2}$ is a better probe for studying the hydrodynamic properties of the quark-gluon plasma. 

\end{abstract}
\pacs{25.75.-q, 25.75.Ld}
\maketitle

\section{Introduction}

One of the primary goals of relativistic heavy ion collisions is to create the quark-gluon plasma (QGP)  where quantum chromodynamics
(QCD) can be studied under the extreme conditions of high temperature and energy density~\cite{Arsene:2004fa,Back:2004je,Adams:2005dq,Adcox:2004mh,Muller:2012zq}. 
In the hydrodynamics picture, the high pressure buildup in the center of the collision zone drives the system to expand at relativistic speed, generating a large collective flow. The collectivity is sensitive to the pressure gradient and therefore regarded as a valuable probe of the pressure and energy density of the transient QGP. 
Of particular interest are non-central collisions where the overlap collision zone is anisotropic in the transverse plane perpendicular to beam. 
The pressure gradient would generate an anisotropic expansion and final-state anisotropic flow in momentum space, whose leading term is elliptical~\cite{Ollitrault:1992bk}. This large elliptic anisotropy ($v_{2}$) has been observed in experiments and is regarded as a strong evidence for the formation of the QGP~\cite{Arsene:2004fa,Back:2004je,Adams:2005dq,Adcox:2004mh,Muller:2012zq}. 

Both hydrodynamics and transport models can describe the measured large $v_{2}$~\cite{Heinz:2013th}.
Hydrodynamics is a macroscopic approach to describe the dynamical evolution of a heavy-ion collision~\cite{Heinz:2013th}.
The fact that hydrodynamics with a small viscosity to entropy density ratio ($\eta / s $) can well describe the experimental data suggests that the collision system is strongly interacting and the created QGP is a nearly perfect fluid (sQGP)~\cite{Gyulassy:2004zy}. 
On the other hand, A Multi-Phase Transport (AMPT)~\cite{Lin:2001zk,Lin:2004en} description of data using perturbative QCD motivated $\sigma$ of 3 mb requires the string melting mechanism. 
Recently observed large $v_{2}$ in small systems of high multiplicity $p+p$ and p+Pb collisions at the Large Hadron Collider (LHC) and $d+Au$ collisions at 
the Relativistic Heavy Ion Collider (RHIC) can also be described well by AMPT with string melting~\cite{Bzdak:2014dia}. 

A recent study within the AMPT framework indicates that at RHIC energies the large $v_{2}$ in the model comes mainly from the escape mechanism~\cite{He:2015hfa,Lin:2015ucn}, where the partons have a larger probability to escape along the shorter axis of the overlap volume and the hydrodynamic flow plays only a minor role in AMPT. 
The escape mechanism would naturally explain the similar azimuthal anisotropies in heavy ion and small systems~\cite{He:2015hfa,Lin:2015ucn,Jaiswal:2017dxp,Kurkela:2018qeb}. It is further shown by AMPT that the main origin of the mass splitting of identified hadron elliptic anisotropies is hadronic scatterings~\cite{Li:2016ubw,Li:2016flp,Li:2016hbf}, thus the mass ordering is not a unique signature of hydrodynamics as naively perceived~\cite{Heinz:2013th}. In this paper, we extend those previous studies to charm quarks at both RHIC and LHC energies. We address the particular question of whether the escape mechanism is also responsible for the development of charm $v_{2}$.

\section{Model and Analysis method}

The string melting version of the AMPT model~\cite{Lin:2001zk,Lin:2004en} consists of four components: fluctuating initial conditions, parton elastic scatterings, quark coalescence for hadronization, and hadronic interactions. We employ the same string melting version of AMPT, v2.26t5, as in our earlier studies~\cite{He:2015hfa,Lin:2015ucn,Li:2016ubw,Li:2016flp,Li:2016hbf}. The initial partons information is provided by HIJING, and then string melting liberates strings into a larger number of quarks and antiquarks. 
The two-body parton-parton elastic scatterings are modeled by Zhang's Parton Cascade (ZPC)~\cite{Zhang:1997ej}. 
We use a Debye-screened differential cross-section $d\sigma/dt\propto\alpha_s^2/(t-\mu_D^2)^2$~\cite{Lin:2004en}, with strong coupling constant $\alpha_s=0.33$ and Debye screening mass $\mu_D=2.265$/fm (corresponding to $\sigma=3$ mb) for all AMPT simulations in this work. Once partons stop interacting, a simple quark coalescence model is applied to combine two (three) nearest partons into a meson (baryon or antibaryon). The subsequent hadronic interactions are described by an extended ART model~\cite{Lin:2004en}. In this study, however, we only study parton-level observables.

We simulate three collision systems: {Au+Au} collisions with {$b=6.6$-8.1~fm} 
at the nucleon-nucleon center-of-mass energy ${\sqrt{s_{_{\rm NN}}}}=200$~GeV, {p+Pb} collisions with {$b=0$~fm} at ${\sqrt{s_{_{\rm NN}}}}=5$~TeV, and {Pb+Pb} collisions with {$b=8$~fm} at ${\sqrt{s_{_{\rm NN}}}}=2.76$~TeV. AMPT uses the same light quark masses as PYTHIA~\cite{Sjostrand:2000wi}: $m_{u}$ =5.6, $m_{d}$=9.9, $m_{s}$=199 MeV/$c^{2}$, 
and we use $m_{c}$=1.2 GeV/$c^{2}$ for charm quark in this study. For simplicity, we assume the same cross section for all flavors; we also require each event to have at least one initial $c-\bar{c}$ pair. We analyze the momentum-space azimuthal anisotropy of partons of different flavors in the final state before hadronization. The momentum anisotropy is characterized by the Fourier coefficients according to $v_{2}=\langle \cos 2(\phi -\psi_2^{(r)}) \rangle$, where $\phi$ is the azimuthal angle of a parton in momentum space~\cite{Voloshin:1994mz} and $\psi_2^{(r)}$ is the harmonic plane of each event from its initial spatial configuration of all partons~\cite{Ollitrault:1993ba}. All results shown in this study are for partons within the pseudo-rapidity window $|\eta|<1$.

\section{Results}

\begin{figure}
\centering
\begin{minipage}[c]{0.4\textwidth}
\centering
\includegraphics[height=4.5cm,width=7.5cm]{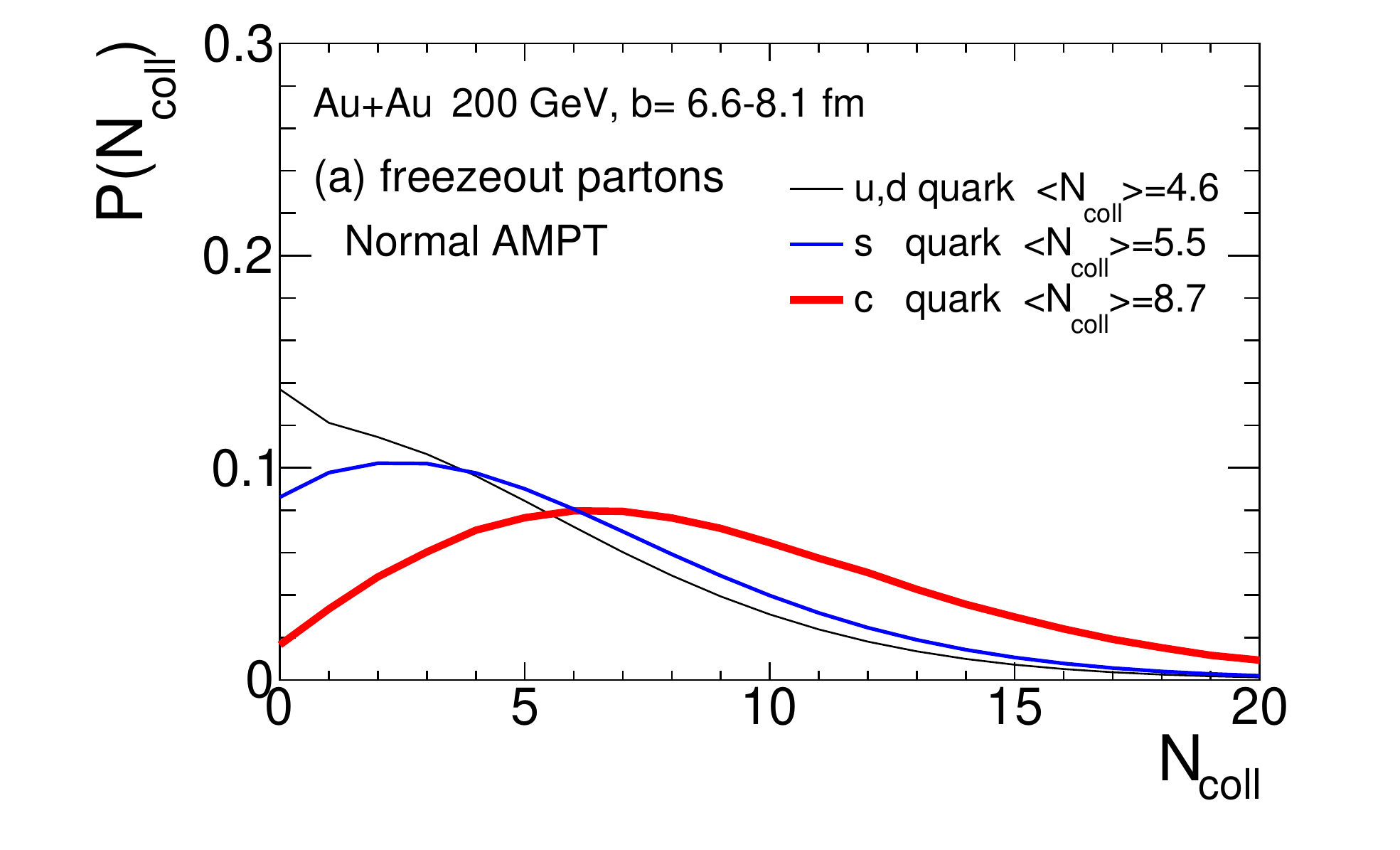}
\end{minipage}
\begin{minipage}[c]{0.5\textwidth}
\centering
\includegraphics[height=4.5cm,width=7.5cm]{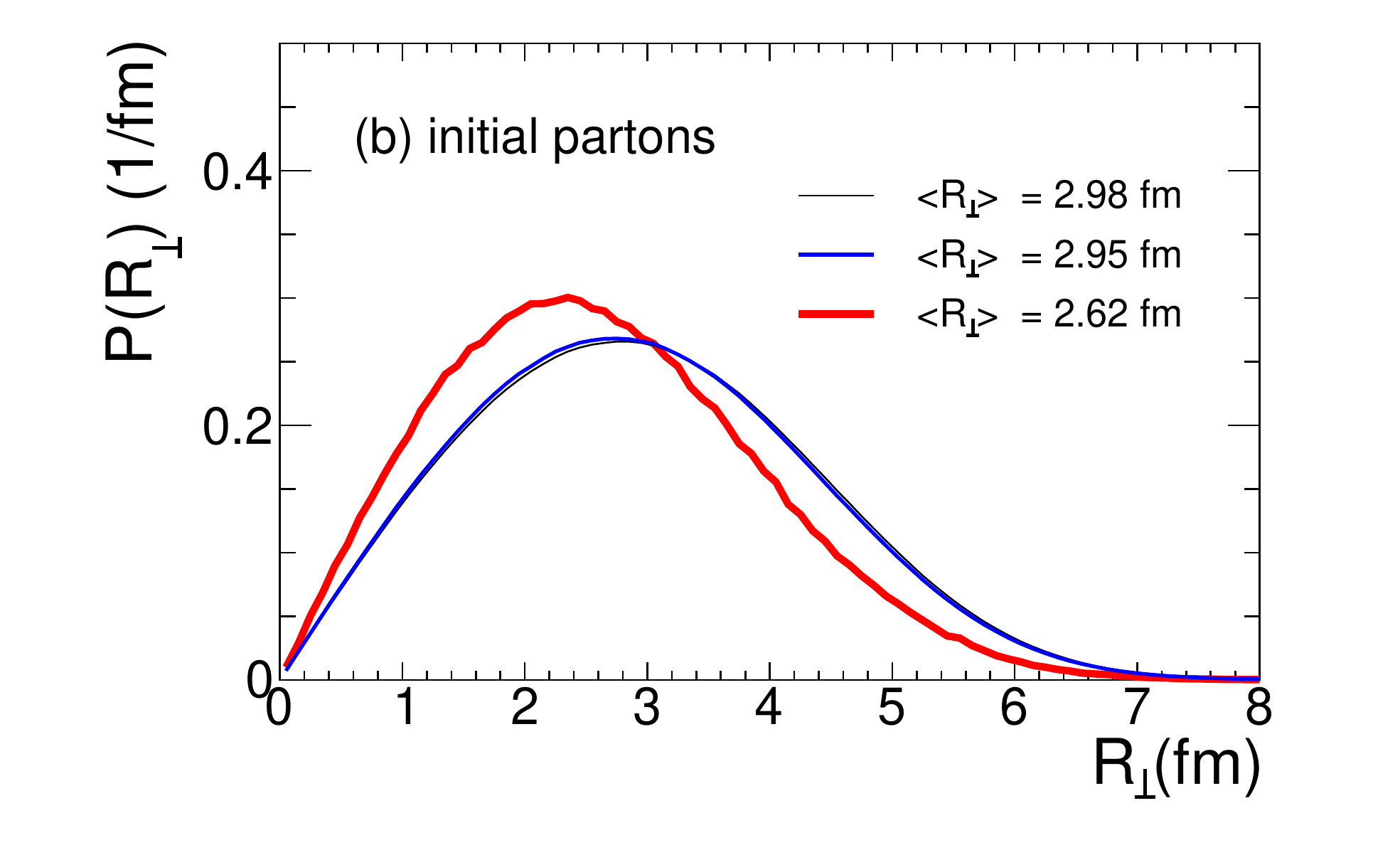}
\end{minipage}
\caption{ AMPT results for {Au+Au} collisions at $\snn=200$A~GeV with impact parameter {$b=6.6$-8.1~fm}: (a) normalized $N_{\rm coll}$-distributions of different quark flavors: light quarks (thin), strange quarks (medium), and charm quarks (thick); (b) normalized probability distributions of the initial transverse radius $R_\perp$ for different quark flavors.  }
\label{fig:pnc}
\end{figure}

We trace the whole collision history of quarks of different flavors including the number of collisions ($N_{\rm coll}$) that a parton suffers with other partons. Fig.~\ref{fig:pnc}(a) shows the normalized probability distributions of each quark flavor that freezes out after $N_{\rm coll}$ collisions in Au+Au collisions. The average number of collisions is flavor (or mass) dependent and charm quarks suffer on average more collisions than light quarks. Such a flavor dependence is consistent with the expectation that the heavier charm quarks are produced by hard scatterings at earlier times from perturbative QCD processes. The flavor dependence is also related to the initial production; for example, Fig.~\ref{fig:pnc}(b) shows that charm quarks are produced in the more inner region of the overlap volume than light quarks. This is partly responsible for the larger charm quark $N_{\rm coll}$ than light 
quark's as shown in Fig.~\ref{fig:pnc}(a). We also find that the features shown in Fig.~\ref{fig:pnc} are qualitatively the same for the p+Pb and Pb+Pb collision systems~\cite{Lin:2018hbf}. 

\begin{figure}[hbt]
  \begin{center}
    \includegraphics[width=1.\columnwidth]{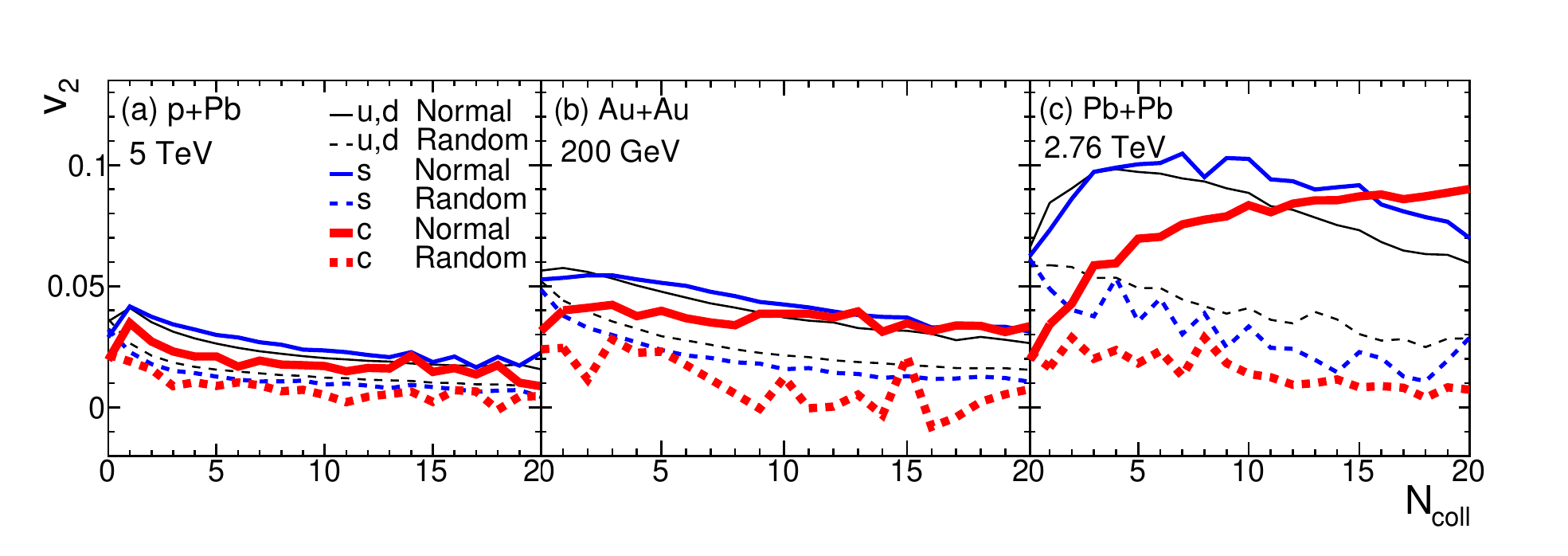}
    \caption{Freezeout partons' $v_{2}$ within $|\eta|<1$ 
as a function of $N_{\rm coll}$ for light (thin), strange (medium), and charm quarks (thick) in (a) {p+Pb} collisions with {$b=0$~fm} at $\snn=5$A~TeV, (b) {Au+Au} collisions with {$b=6.6$-8.1~fm} at $200$A~GeV, and (c) {Pb+Pb} collisions with {$b=8$~fm} at $2.76$A~TeV in normal AMPT (solid curves) and  $\phi$-randomized AMPT (dashed curves).}
    \label{fig:udsc}
  \end{center}
\end{figure}

Our previous studies~\cite{He:2015hfa,Lin:2015ucn} have shown that the overall quark $v_{2}$ at RHIC energies 
comes mainly from the anisotropic escape mechanism. The question is whether or not this is also true for charm $v_{2}$. To address this question, we compare in Fig.~\ref{fig:udsc} the $v_{2}$ of freezeout partons of different flavors in both the normal AMPT calculation and the azimuth-randomized AMPT calculation. Since $v_{2}$ values for quarks and antiquarks of the same flavor are practically identical at these high energies, we combine the quarks and antiquarks of the same flavor in the $v_{2}$ plots. 
In the normal calculations, there is a mass ordering in $v_{2}$ at the same $N_{\rm coll}$: 
the light quark $v_{2}$ is larger than the charm quark $v_{2}$ at small $N_{\rm coll}$, while the ordering is reversed at large $N_{\rm coll}$ in collisions of large systems. 
It suggests that charm quarks need more scatterings to accumulate large $v_{2}$ while the light quarks need only a few collisions. 
In the randomized case, the parton azimuthal angles are randomized after each collision and hence their $v_{2}$ only comes from the anisotropic escape mechanism \cite{He:2015hfa,Lin:2015ucn}. 
The freezeout partons still have positive $v_{2}$ but the values are reduced from those in the normal case; this is because of no contribution from the anisotropic collective flow.  
The results shown in Fig.~\ref{fig:udsc} indicate that the escape mechanism is qualitatively at work for both charm $v_{2}$ and light quark (u,d,s) $v_{2}$. 
Quantitatively, however, we see that the hydrodynamics-type flow has a larger contribution to charm $v_{2}$ than the light quark $v_{2}$ at large $N_{\rm coll}$. 
Note that the significant contribution of hydrodynamic flow to the charm $v_{2}$ has been observed in a linearized Boltzmann transport model of heavy quarks coupled with hydrodynamical background~\cite{Cao:2016gvr}.

\begin{figure}[tbph]
\centering
 \includegraphics[width=1.\columnwidth]{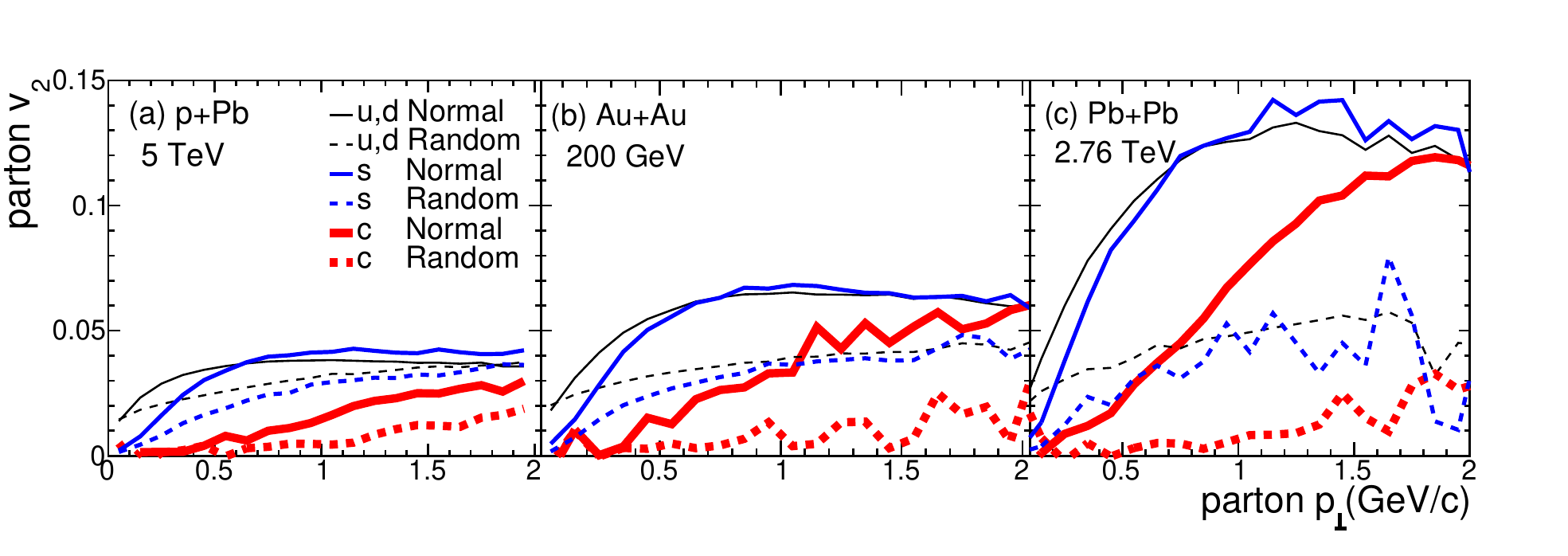}
 \caption{(Color online) {Parton $v_{2}$ as a function of $p_\perp$ at freezeout for different quark flavors: light quarks (thin), strange quarks (medium), and charm quarks (thick) in normal (solid curves) and $\phi$-randomized (dashed curves) AMPT for three different collision systems.
}
 \label{fig:udsc_pt}}
\end{figure}

Fig.~\ref{fig:udsc_pt} shows the final quark $v_{2}$ as a function of $p_\perp$ for different quark flavors at the final state in normal and azimuth-randomized AMPT calculations.
Like Fig.~\ref{fig:udsc}, the escape mechanism contribution as shown by the azimuth-randomized AMPT results is finite~\cite{He:2015hfa} for each quark flavor. The mass splitting of quark $v_{2}$ is present at low $\pt$ for both $\phi$-randomized and normal AMPT. 
At high $\pt$, however, their $v_{2}$'s from normal AMPT simulations approach each other as expected because the mass difference becomes less important. 
We also see that the escape mechanism contributes less to the charm $v_{2}$ than to the light quark $v_{2}$, therefore the hydrodynamic flow contributes more to the charm $v_{2}$ than to the light quark $v_{2}$; this is especially clear for {Au+Au} and {Pb+Pb} collisions. 
Since the escape mechanism dominates more the $v_{2}$ of small systems~\cite{He:2015hfa,Lin:2015ucn}, 
the percentage contribution from the hydrodynamic flow to the charm $v_{2}$ is larger in {Pb+Pb} collisions than {p+Pb} collisions. 

Table~\ref{table:rnp} lists the values of $\langle N_{\rm coll} \rangle$, $\langle v_{2} \rangle$,  and the ratios of $\langle v_{2} \rangle$ from azimuth-randomized AMPT to that from normal AMPT in {Au+Au}, {p+Pb} and {Pb+Pb} collisions. The $\langle N_{\rm coll} \rangle$ value of freezeout partons of a given flavor increases with the collision system size and beam energy as expected. Note that we also find that the $\langle N_{\rm coll} \rangle$ values are larger in the randomized case compared to normal AMPT; this is because the randomization tends to destroy the preferred outward direction of partons. 
We see in Table~\ref{table:rnp} that the $\langle v_{2} \rangle$ values of freezeout partons of different flavors are not very different in normal AMPT calculations. 
However, the $\langle v_{2} \rangle$ values of freezeout partons in the azimuth-randomized AMPT 
strongly depend on the quark flavor: the light quark $\langle v_{2} \rangle$ is higher than charm quark  $\langle v_{2} \rangle$ for the same collision system.

\begin{table*}[!htp]
\begin{center}
\begin{tabular}{c|ccc|ccc|cccl}\hline
& \multicolumn{3}{c|}{pPb ({$b=0$~fm})}& \multicolumn{3}{c|}{AuAu ({$b=6.6$-8.1~fm})} & \multicolumn{3}{c}{PbPb ({$b=8$~fm})} \\\hline
 Quark flavor & {u,d} & {s} & {c} & {u,d} & {s} & {c} & {u,d} & {s} & {c} \\\hline
$\mean{\Ncoll}$ & 2.02 & 2.54 & 4.23 & 4.58 & 5.45 & 8.68 & 9.82 & 11.14 & 15.48\\
$\mean{v_2}_{\rm Random}$ & 2.39\% & 1.89\% & 1.21\% & 2.93\% & 2.27\% & 0.85\% & 3.21\% & 2.23\% & 0.67\%\\
$\mean{v_2}_{\rm Normal}$ & 3.28\% & 3.20\% & 2.14\% & 4.47\% & 4.78\% & 3.89\% & 7.56\% & 8.42\% & 7.92\%\\
$\mean{v_2}_{\rm Random}/\mean{v_2}_{\rm Normal}$ & 73\% & 59\% & 57\% & 66\% & 47\% & 22\% & 43\% & 27\% & 8.5\%\\\hline
\end{tabular}
\caption{\label{table:rnp} $\langle N_{\rm coll} \rangle$ , $\mean{v_2}_{\rm Random}$, $\mean{v_2}_{\rm Normal}$ and the ratio of $\langle v_{2} \rangle$ from $\phi$-randomized AMPT over that from normal AMPT for final partons of different flavors within $|\eta|<1$ in three collision systems.}
\end{center}
\end{table*}

The ratio of $\langle v_{2} \rangle$ from $\phi$-randomized AMPT to that from normal AMPT represents the fraction of $v_{2}$ that comes from the escape mechanism. Table~\ref{table:rnp} shows that the escape mechanism contribution to final quark $\langle v_{2} \rangle$ is smaller for charm quarks. Consequently, the hydrodynamic contribution to the final quark $\langle v_{2} \rangle$ is more important for charm quarks. 
This result suggests that the charm $v_{2}$ better reflects the hydrodynamic properties of the quark-gluon plasma, especially for large systems at high energies. This is consistent with several recent findings~\cite{Esha:2016svw,Greco:2017rro}.

\section{Understanding with a toy model}

\begin{figure}
\centering
\begin{minipage}[c]{0.4\textwidth}
\centering
\includegraphics[height=4.5cm,width=7.5cm]{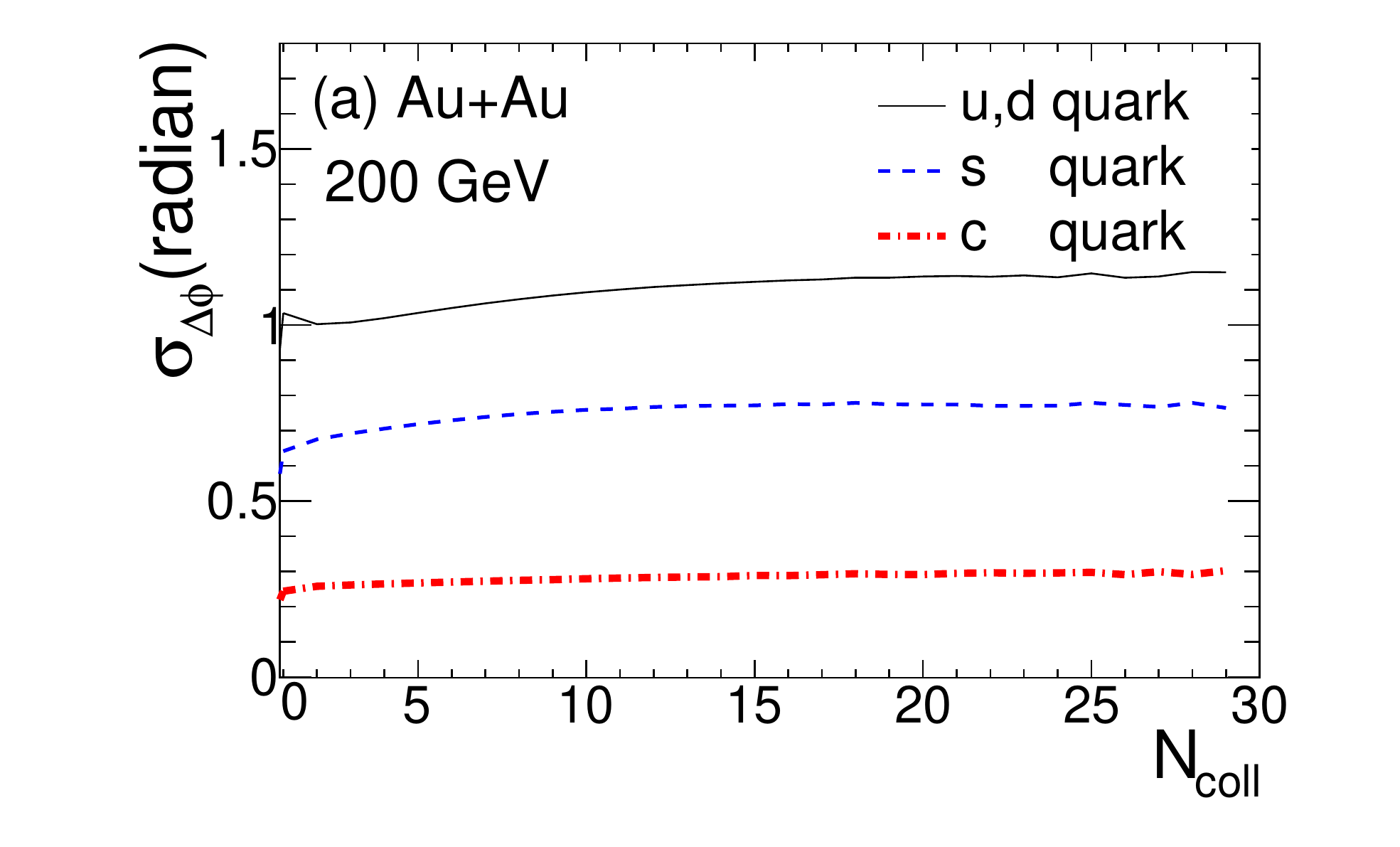}
\end{minipage}%
\begin{minipage}[c]{0.5\textwidth}
\centering
\includegraphics[height=4.5cm,width=7.5cm]{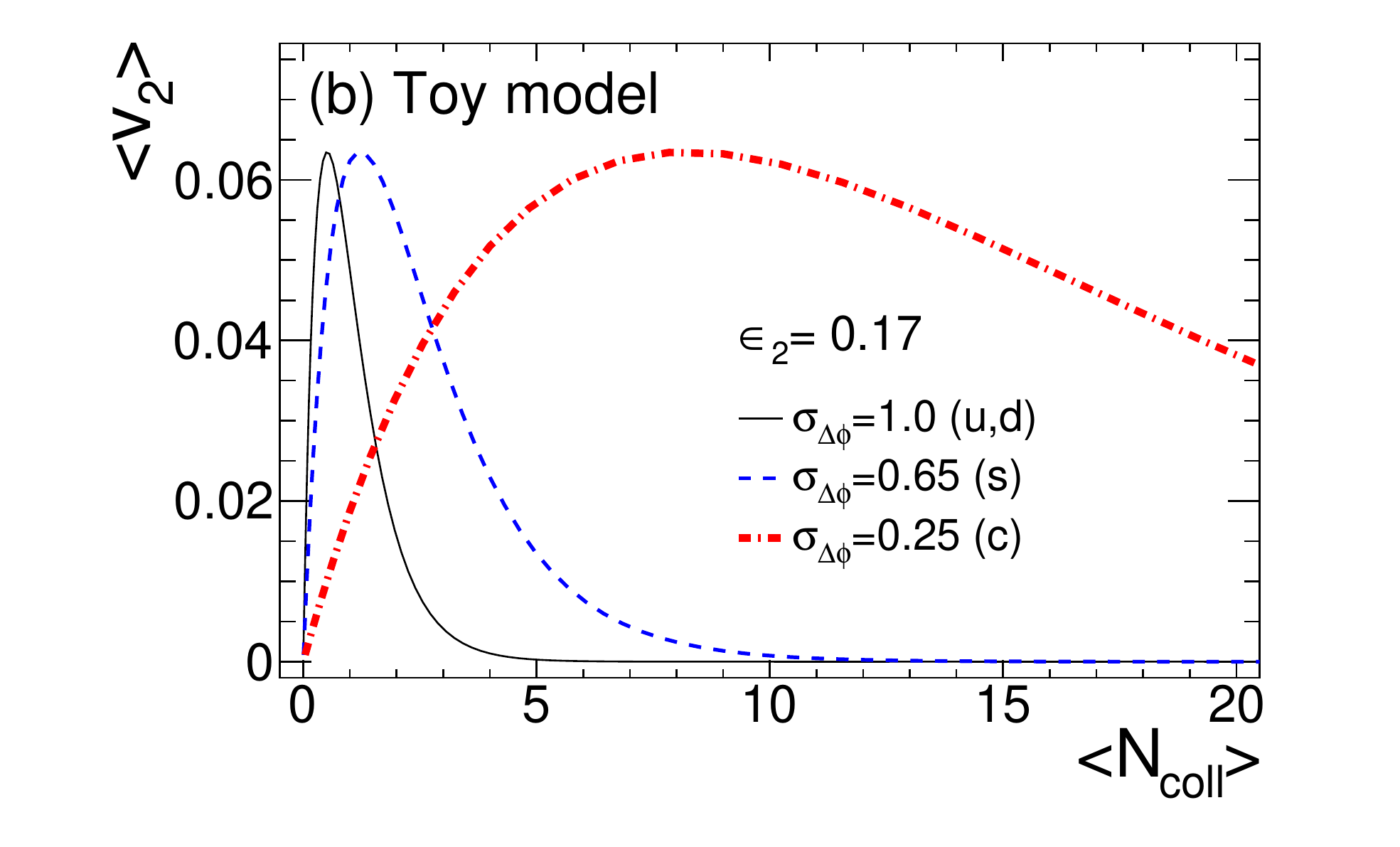}
\end{minipage}
\caption{(a) The rms change of azimuth due to the $N_{\rm coll}$-th collision for different quark flavors in normal AMPT simulations in {Au+Au} collisions. The black curves are for light quarks, the blue curves for strange, and red curves for charm quarks. (b) The freezeout partons $\langle v_{2} \rangle$ as a function of $\langle N_{\rm coll} \rangle$ for light, strange, and charm quarks from a simple analytical calculation for a transverse geometry of a given $\epsilon_{2}=0.17$, with arbitary size nucleus-nucleus collision quantified by the variable $\langle N_{\rm coll} \rangle$.}
\label{fig:mc}
\end{figure}

For parton scatterings, a main difference between light and heavy quarks is their masses. In the AMPT model used for the current study, the mass difference leads to a difference in the average deflection angle in a 2$\rightarrow$2 scattering. Figure~\ref{fig:mc}(a) shows the root-mean-square (rms) change of the azimuth angle ($\sigma_{\Delta\phi}$) as a function of $N_{\rm coll}$ for different quark flavors in {Au+Au} collisions. The results are similar for {p+Pb} and {Pb+Pb} collisions~\cite{Lin:2018hbf}. The rms change of the azimuthal angle is approximately $\sigma_{\Delta\phi}$=1.0, 0.65, 0.25 for light, strange, and charm quarks, respectively; so it is much smaller for heavier quarks~\cite{Svetitsky:1987gq}. 

For better understanding, we now consider a toy model where partons start from the center $(x, y) = (0, 0)$ and propagate out to the boundary of an ellipse. We take the eccentricity of the elliptical area to be $\epsilon_{2}=0.17$, corresponding to semi-central Au+Au collisions. The number of collisions a parton suffers, assuming
straightline propagation, can be written as
\begin{eqnarray}
N_{\rm coll}(\phi_{i})=\langle N_{\rm coll} \rangle \left ( 1-2\epsilon_{2} \cos 2\phi_{i} \right ),
\end{eqnarray}
where $\phi_{i}$ is the initial azimuthal angle. Note that $\langle N_{\rm coll} \rangle$ here is a measure of the system size. For example, $\langle N_{\rm coll} \rangle =5$ corresponds to a mid-central  {Au+Au} collisions, and $\langle N_{\rm coll} \rangle =20$ would correspond to a mid-central collision between two hypothetical large nuclei. The cumulative deflection in the azimuth angle 
after the parton leaves the elliptical area, assuming straightline propagation, would be Gaussian distributed with the width of $ \sigma_{\Delta\phi} \sqrt{N_{\rm coll}(\phi_{i})}$.

Then the parton average elliptic flow $\langle v_{2} \rangle$ can be calculated as
\begin{eqnarray}
\langle v_{2} \rangle=\frac{1}{(2\pi)^{3/2}\sigma_{\Delta\phi} \sqrt{\langle N_{\rm coll} \rangle}}\int \frac{\cos2(\phi_i+\delta \phi)}{\sqrt{1-2\epsilon_{2}\cos2\phi_i}}
\exp{ \left( - \frac{ \delta \phi^{2}}{2\sigma^{2}_{\Delta\phi} \langle N_{\rm coll} \rangle (1-2\epsilon_{2} \cos2\phi_i)}\right )} d\phi_i d\delta \phi.
\end{eqnarray}
The results are shown in Fig.~\ref{fig:mc}(b). The $\langle v_{2} \rangle$ of light quarks is larger than charm quarks at small $\langle N_{\rm coll} \rangle$ but is the opposite at large $\langle N_{\rm coll} \rangle$. Partons along the longer $y$-axis suffer more collisions than those along the $x$-axis, and each collision deflects the parton from its original direction to a range of directions. As a result, more $y$-going partons will be deflected towards an isotropic distribution. This results in a positive $v_{2}$, and this is a generic mechanism for the $v_{2}$ generation in any cascade model. Since $\sigma_{\Delta\phi}$ is large for light quarks, a small number of collisions is already strong enough to reshuffle the $\phi$ directions to produce a large $v_{2}$. However, also because $\sigma_{\Delta\phi}$ is large, light quarks easily forget their original direction, so the light quark $v_{2}$ quickly drops to zero at modest $\langle N_{\rm coll} \rangle$ in this toy model. 
For charm quarks, on the other hand, it takes many collisions to build up a sizable
$v_{2}$ because each collision only deflects the charm quark direction a little. 
In addition, since $\langle v_{2} \rangle$ only depends on the variable 
$\sigma_{\Delta\phi} \sqrt{ \langle N_{\rm coll} \rangle}$, the average elliptic flow 
$\langle v_{2} \rangle$ has the same peak value for all flavors, while the peak occurs at a larger $\langle N_{\rm coll}\rangle$ value for heavier quarks. 
We note that this toy model does not have much dynamics and thus does not capture certain important features of dynamical models, such as the finite asymptotic $v_{2}$ after infinite number of collisions. Instead, the $v_{2}$ from the toy model basically represents the $v_{2}$ generated by the escape mechanism. In the toy model partons of all flavors will be randomized and reach zero $\langle v_{2} \rangle$ after a large number of collisions, where the hydrodynamic collective flow would be the dominant source of anisotropic flows~\cite{He:2015hfa,Lin:2015ucn}. Because our toy model assumes straightline propagation, it cannot be trusted at large $\langle N_{\rm coll} \rangle$ when $\sigma_{\Delta\phi} \sqrt{ \langle N_{\rm coll} \rangle}$ is of order 1. However, our toy model calculation helps to illustrate the importance of the average scattering deflection angle, thus also the importance of quark mass, for the generation of $v_{2}$ by parton scatterings. 

It is commonly expected that transport with large number of scatterings would approach hydrodynamic limit. If hydrodynamic evolution proceeds with a quasiparticle picture, then it is conceivable that light quarks do not accumulate $v_{2}$ efficiently because they can easily change directions, and heavy flavors attain better the accumulated $v_{2}$ to the final state.
In other words, heavy flavor final-state flow would appear more hydrodynamic than the light flavor flow, even though light flavor approaches hydrodynamic equilibrium earlier in the collision. This would imply that light quarks can partially lose their evolution memory because of thermalization, and this includes also their $v_{2}$ which has been generally perceived to accumulate to the final state. 

\section{Summary}

Recent transport model studies have shown that the overall azimuthal anisotropies $v_{n}$ in heavy ion collisions come mainly from the anisotropic escape probability in 
heavy ion collisions at RHIC energies and in small collision systems. 
Here we have extended those previous studies to quarks of different flavors, particularly  
the charm quarks. We find that the common escape mechanism is at work not only for light quark but also for strange and charm
quark $v_{2}$, where it dominates the generation of $v_{2}$ for small-enough systems but would become insignificant for large-enough systems. However, we find that charm $v_{2}$ has a larger fraction coming from the hydrodynamic collective flow (and thus less coming from the escape mechanism) than the light quark $v_{2}$, in contrast to naive expectations. We further find that this is closely related to the mass dependence of the average scattering angles. Light quarks have large deflection angles and thus cannot efficiently accumulate their  $v_{2}$ (their final state  $v_{2}$ comes largely from the escape mechanism), while heavy quarks deflect at smaller angles and thus need more scatterings with the light-quark-dominated medium in order to accumulate a significant $v_{2}$. We conclude that the charm $v_{2}$ is more hydrodynamic than the light quark $v_{2}$, therefore it is a better probe of the hydrodynamic properties of QGP.

{\em Acknowledgments.}
This work is supported in part by Hubei Province Key Laboratory of Systems Science in Metallurgical Process (Wuhan University of Science and Technology) No.~Y201710,
the National Natural Science Foundation of China under Grants Nos.~11628508, 11647306, and 11747312, and US~Department of Energy Grant No.~DE-SC0012910.
HL also acknowledges financial support from the China Scholarship Council.

\bibliography{./ref}\end{document}